\documentclass[preprint,12pt]{elsarticle}
\usepackage{graphicx}
\usepackage{amssymb}
\journal{
Physica A: Statistical Mechanics and its Applications}
\begin{document}
\begin{frontmatter}
\title{Detecting spatial homogeneity in the World Trade Web with
Detrended Fluctuation Analysis}
\author[label1,label2]{Riccardo Chiarucci}
\author[label1,label3,label4]{Franco Ruzzenenti\corref{cor1}}
\author[label1,label2]{Maria I. Loffredo}
\address[label1]{CSC$-$Complex Systems Community, Siena}
\address[label2]{Department of Information Engineering and
Mathematics, University of Siena, Italy}
\address[label3]{Department of Economics and Statistics, University of
Siena, Italy}
\address[label4]{Department of Biotechnology, Chemistry and Pharmacy,
University of Siena, Italy}
\cortext[cor1]{ruzzenenti@unisi.it}
\cortext[cor1]{The first two authors equally contributed to the work}
\begin{abstract}
In a spatially embedded network, that is a network where nodes can be
uniquely determined in a system of coordinates, links' weights might
be affected by metric distances coupling every pair of nodes (dyads).
In order to assess to what extent metric distances affect
relationships (link's weights) in a spatially embedded network, we
propose a methodology based on DFA (Detrended Fluctuation Analysis).
DFA is a well developed methodology to evaluate autocorrelations and
estimate long-range behaviour in time series. We argue it can be
further extended to spatially ordered series in order to assess
autocorrelations in values. A scaling exponent of $0.5$ (uncorrelated
data) would thereby signal a perfect homogeneous space embedding the
network. We apply the proposed methodology to the World Trade Web
(WTW) during the years 1949-2000 and we find, in some contrast with
predictions of gravity models, a declining influence of distances on
trading relationships.
\end{abstract}
\begin{keyword}
DFA analysis \sep distance puzzle \sep spatially embedded networks
\sep spatial homogeneity \sep exponential random graphs
\end{keyword}
\end{frontmatter}
\section{Introduction}
\label{intro}
In studying dynamical systems, homogeneity of space is a fundamental assumption of particle interactions. As is well known, invariance under translation or rotation or, alternatively, breaking of symmetries in presence of a force field, play a crucial role in physical systems. In these cases distances certainly do count. Moreover the question of defining and assessing the role of distances in shaping social and economic complex systems - like communities, cities and
whole economies - has always been addressed by social sciences,
economics, human geography and more recently by network theory of
spatial networks  \cite{Barthelemy} .
The most remarkable example of such system, where the distance
factor has undergone a long-standing investigation, is given by the
trading relationships between countries. So far as the early 1960s
gravity models were proposed to explain bilateral trades flows
\cite{Tinbergen}.
Gravity models functionally resemble the two bodies' interaction
gravity formula, where the mass of the particle is replaced by the GDP
(Gross Domestic Product) or by any other measure of the size of the
economy. Gravity models proved to work pretty well in predicting the
intensity of trading flows between any pair of countries, given the
size of the two economies and the geographic distance. Nevertheless,
they fail to predict the \emph{existence} of the flow. In other words,
they are not able to explain any {\it zero-intensity} flow between any
pair of countries with a {\it non-zero} GDP  \cite{DuenasFagiolo}.
Therefore, according to gravity models, any country should trade with
any other, or, in a mathematical jargon, the network of the world-wide
trading relationships should be a \emph{full graph}. Conversely, we know that the World Trade Web (WTW) has a non trivial topological structure, resembling those of many complex networks, in society and nature
\cite{GarlaschelliLoffredo2004a,GarlaschelliLoffredo2004b, SquartiniFagioloGarlaschelli}.
A second major shortcoming of gravity models concerns the consistency
and robustness of the distance parameter estimations along time.
Despite being very recent, the so called \emph{distance puzzle} is a
very debated riddle in the economic literature \cite{AndersonYotov,
BuchKleinertToubal} . According to standard gravity models, based on
log-linear regressions, the parameter defining the weight of distances
has always been considered constant, indicating that, contrary to
expectations, transport costs, barrier removal and market integration
- among the many factors favoring globalization - never affected the
role of distances in shaping the WTW \cite{RuzzenentiBasosi}.
In this paper we propose to test the relevance of distances in the
trading network by using a procedure based on the detrended
fluctuation analysis.
Even if this methodology is typically applied to analyze
autocorrelation properties and long-range behavior of data which are
temporally ordered, we propose to extend it in order to verify the
homogeneity of the space in which a network is embedded and the role
of the distances in it.
This goal can be achieved only after a proper procedure of spatial
reordering of the data - based on metric distances between vertices -
has been implemented.
Moreover the homogeneity of the space embedding the network can be
considered as the null hypothesis corresponding to absence of
correlations, hypotheses that can in fact be tested by using the DFA.
We apply the proposed procedure to the Trading Network during a
temporal window of fifty years starting from the year 1949. We find -
as a main outcome - a declining influence of distances on trading
relationships.
As a byproduct our results could in part explain the ``distance
puzzle" which is one of the drawback inside the gravity models.
\section{Method}
\label{method}
\subsection{Applying DFA to time series}
\begin{figure}
\includegraphics[scale=0.3]{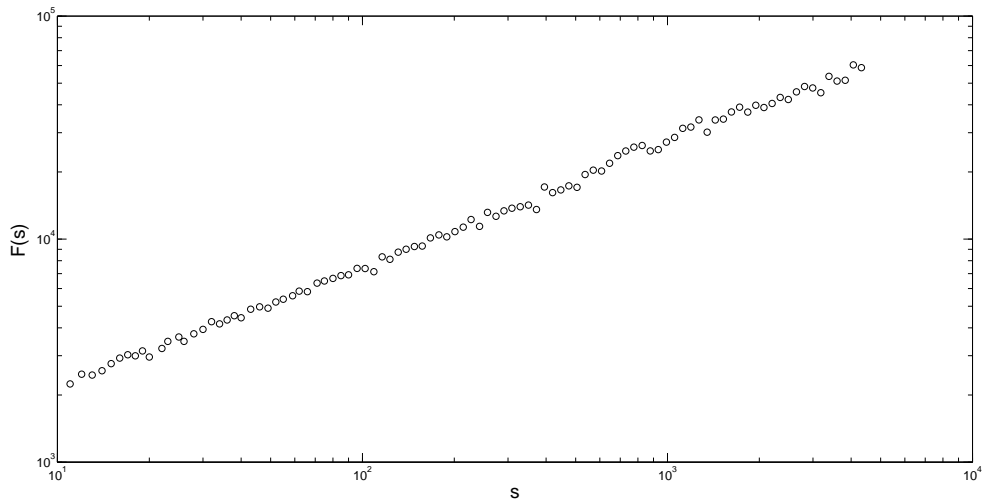}
\label{profile}
\caption{Fluctuation function, WTW, year 2000}
\centering
\end{figure}

DFA has been extensively used to estimate long-range power-law
correlation exponents in noisy signals \cite{Stanley1, Stanley2,
Havlin}. More recently, DFA has been further improved in order to cope with series characterized by high non-linearity and periodicity \cite{Stanley4}.
To illustrate the DFA method, we consider a noisy time
series, $u(i)$, $(i = 1, ..,N )$.
We integrate the time series  u(i) and we get the profile
\begin{equation}
y(j)=\sum_{i}^{j}(u(i) - \langle u\rangle)
\end{equation}
where
\begin{equation}
\langle u\rangle=\frac{1}{N} \sum_{i}^{N}u(i)
\end{equation}
We divide the profile into $N_{s}=N/s$  boxes of equal size s. Since
the length of the series is generally not an integer multiple of time
scale s, a small portion of data at the end of the profile will not be
included in any interval. In order not to overlook this small part we
repeat the entire procedure starting from the end of the series,
obtaining $2N_{s}$ segments. Then in each box, we fit the integrated
time series by using a polynomial function,  $y_{fit}(i)$, which is
called the local trend. For $order-l$ $DFA$ polynomial functions of
order $l$ should be applied for the fitting.
We calculate the local trend for each interval of width $s$ via a
linear fit of data. Indicating with $y^{\nu}_{fit}(i)$ the fit in the
$\nu$-th box we define the detrended profile as
\begin{equation}
Y_{s}(i)=y(i)-y^{\nu}_{fit}(i)
\end{equation}
where
\begin{equation}
(\nu -1)s<i<\nu s
\end{equation}
For each of $2N_{s}$ intervals we calculate the mean square deviation
from the local trend
\begin{equation}
F^{2}_{s}(\nu)=\frac{1}{s} \sum_{i=1}^{s}Y_{s}^{2}[(\nu -1)+i]
\end{equation}
Finally, we calculate the mean on all segments to obtain the
\emph{fluctuation function}
\begin{equation}
F(s)=\sqrt{\frac{1}{2N_{s}}\sum_{\nu =1}^{2N_{s}}F_{s}^{2}(\nu)}
\label{flut}
\end{equation}
The above computation is repeated for box sizes s (different scales)
to provide a relationship
between $F(s)$ and $s$. A power-law relation between $F(s)$ and the
box size $s$ indicates the
presence of scaling: $F(s)\propto s^{\alpha}$. As an example,  Figure  \ref{profile} shows the behavior of the fluctuation
function for the WTW (2000), anticipating the use of the corresponding
time series as obtained through the procedure explained in the next
section..
The parameter $\alpha$,
called the scaling exponent or correlation
exponent (which can be directly related to the Hurst Exponent $H$ if
the series is stationary) gives a measure of the correlation
properties of the signal: if $\alpha = 0.5$, there is no correlation
and the signal is an uncorrelated signal(white noise); if $\alpha <
0.5$, the signal is anticorrelated; if $\alpha > 0.5$, there are
positive correlations in the signal \cite{Hurst, Mandelbrot}.
\subsection{Encoding spatial series}
\begin{figure}[ht]
\includegraphics[scale=0.4]{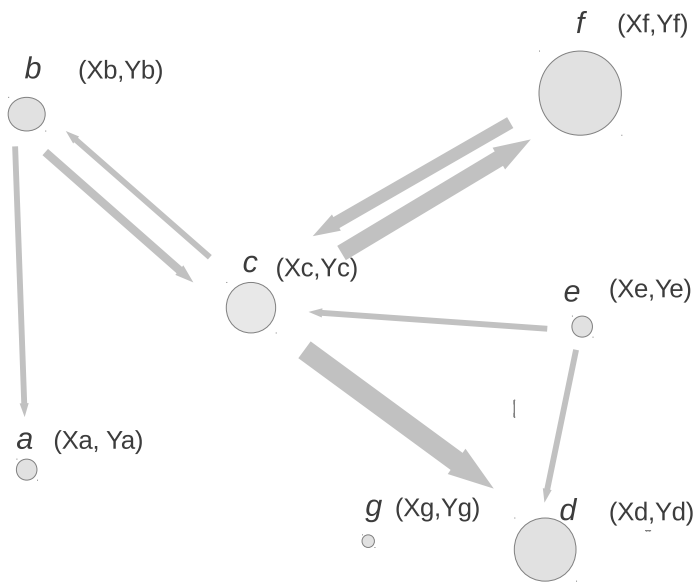}
\caption{Network embedded in a metric space}
\centering
\label{embed1}
\end{figure}

\begin{figure}[ht]
\includegraphics[scale=0.4]{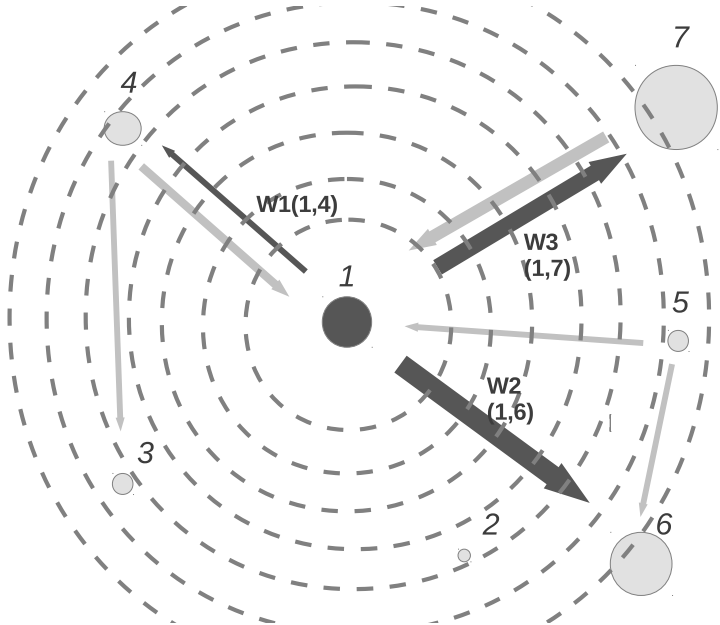}
\caption{Encoding distances from vertex C in graph \ref{embed1}}
\centering
\label{embed2}
\end{figure}

\begin{figure}
\includegraphics[scale=0.3]{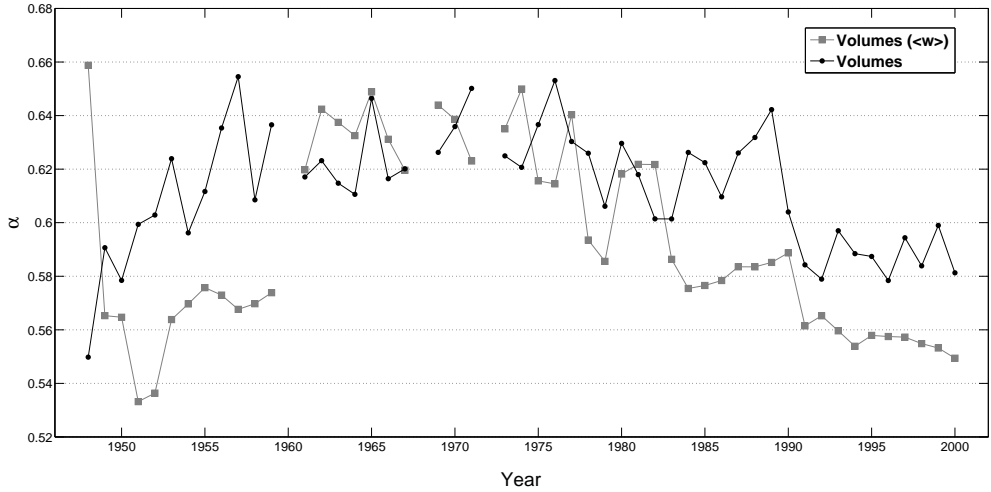}
\caption{Evolution of the $\alpha$ exponent for the WTW, deflated and
non-deflated}
\label{H_non_ordered}
\end{figure}

So far, DFA has been applied mainly to time series, with some
remarkable exceptions. Peng et al. \cite{Stanley3}, applied DFA to
evaluate long-range power-law correlations for DNA sequences
containing noncoding regions. Can we apply DFA to assess autocorrelations in signals which are spatially embedded? By doing that, we aim at assessing the
homegeneity of space for the signal generating system, in the same way
as a $0.5$ exponent would indicate a purely uncorrelated time series.
Indeed, in unidimensional spaces, like the DNA chain, the analogy with
the time series is straightforward and the application of DFA thereof.
What if the space dimension is greater than one?
In what follows, we will explain how we generated a signal spatially
encoded from an embedded network. In a network embedded in a
two-dimensional space, nodes can be univocally defined by a couple of
coordinates (Figure \ref{embed1}).
For every vertex, it is possible to array the outgoing links according
to the distance of the receiving node. In Figure \ref{embed1} we selected node C as the starting node and then we order the outgoing links according the distance of the incoming vertex. The order of the links are explicited in Figure \ref{embed2}.
It is possible to obtain a series of $n-1$ weights (strength of links) for each node. That is to say: we can array the entries of the rows of the weighted matrix according to the distance from the row's first entry. We are
therefore applying a central symmetry scheme, curiously, like in
gravity models. Once engulfed all the nodes, we moved to another node
and started again encoding from the first neighbour onward (Figure
\ref{embed2}). At first, we order the list on $n$ sequences according
to columns'order of the entries in the matrix of the weighted directed network
($a->b->c....->f$ in Figure \ref{embed1}).
While encoding the signal according to nodes' distances, it is possible to overrate the role of space versus the role of topology. Strongly connected nodes (hubs) tend to connect mutually
regardless of the distance (c and f in the example). Similarly,
countries with high GDP values tend to trade more than countries with
a low GDP \cite{GarlaschelliLoffredo2004b}. It was recently proved
that, for the case of the WTW, the topology and the symmetry structure
of the binary network are actually more important in defining trading
relationship than distances
\cite{PiccioloSquartiniRuzzenentiBasosiGarlaschelli}. In the weighted stucture of the WTW too topology plays a fundamental role, both on a local and global scale \cite{SquartinietalBookChapter}.
Therefore, in order to discount topological effects, we deflate the
trading values of the network (volumes), with deflators generated by a
Null Model based on the Exponential Random Graph methodology. We
calculate the expected value of trades over a graph ensemble of
randomized networks that preserved the observed strength sequence \footnote{The Weighted Directed Configuration Model (WDCM) underestimates the reciprocal weighted structure of the WTW and a better null model would probably be the one that preserves the reciprocated strength together with the in- and out- strength. However, this latter model has been recently formalized but not implemented \cite{SquartinietalWrecip}. This is subject for a new research and enhancement of the herein proposed method.} (see
Appendix) and we divide the observed values by the randomly generated
values:
\begin{equation}
w_{ij}^{deflated}=\frac{w_{ij}}{<w_{ij}>}
\label{deflated}
\end{equation}
Figure \ref{H_non_ordered} reports results for both the deflated and
non-deflated series in the WTW.

\begin{figure}
\includegraphics[scale=0.3]{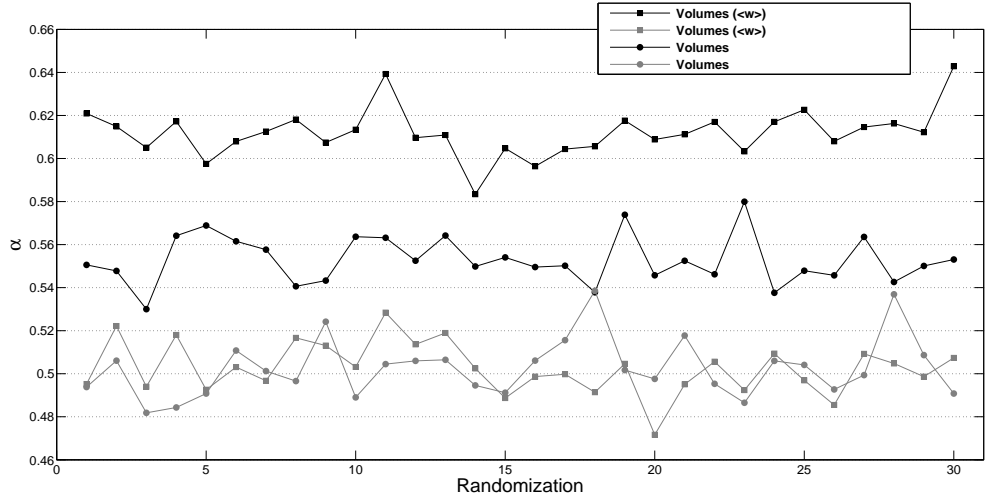}
\caption{Randomization over rows (black) and over rows and columns (grey)}
\label{H_random}
\end{figure}

\begin{figure}
\includegraphics[scale=0.3]{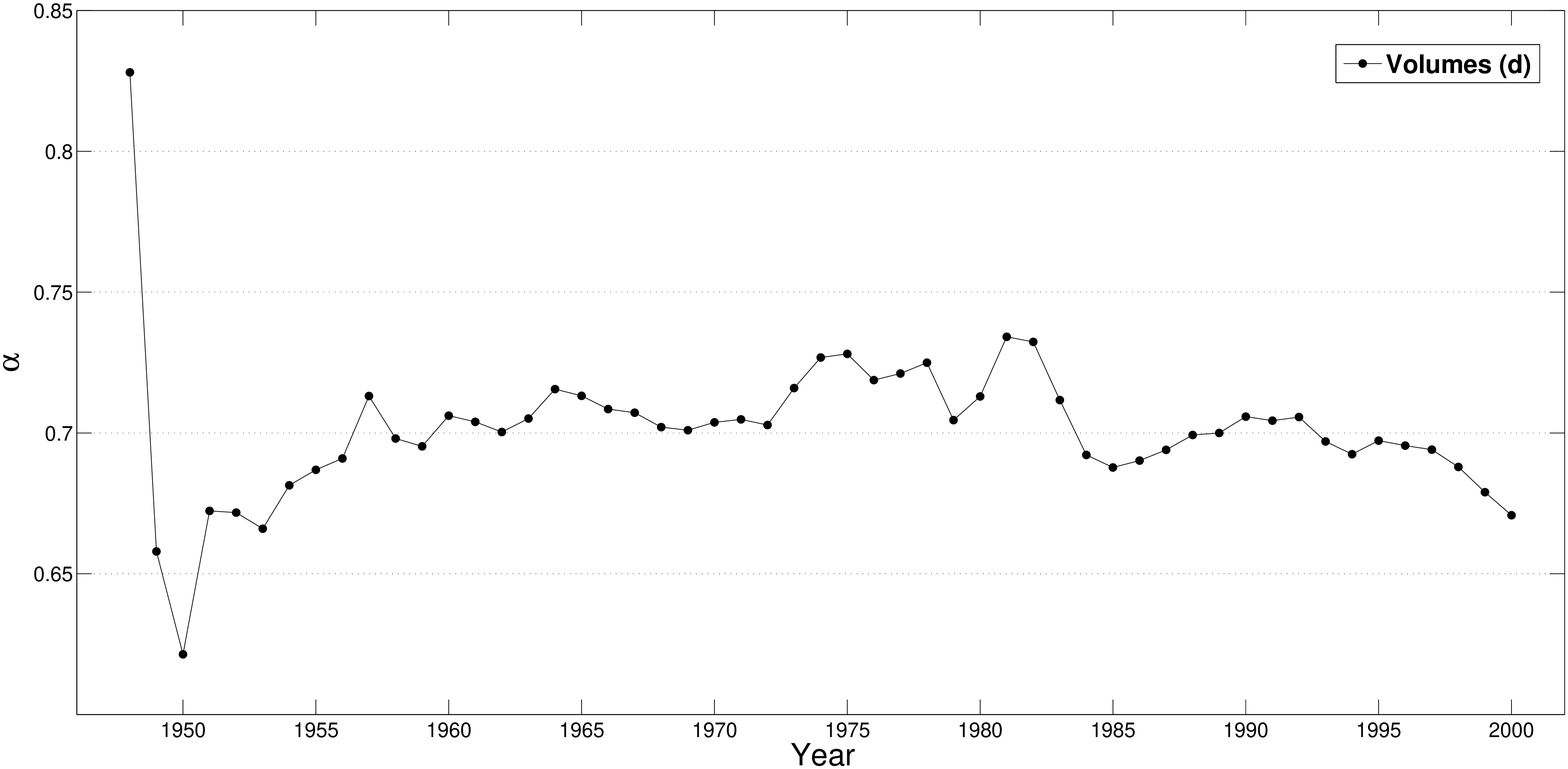}
\caption{Evolution of the $\alpha$ exponent for the WTW, non-deflated,
with nodes ordered according to the distance from a pivot}
\label{H_ordered}
\end{figure}

However, the row's order of the adjacency matrix we have so far chosen
for arraying the $n$ lists of signals must be considered arbitrary. We
asked ourselves whether this arbitrary choice is fatal to DFA by
operating a randomization, for both the volumes' series and the
deflated series, over the row's order (preserving the distance
ordering criteria for every row). We compared those randomizations
with randomizations over all the values (rows and columns). It is
noteworthy that, whereas the $\alpha$ value for the fully randomized
series, as expected, floats around 0.5, with a standard deviation of 0.01, the $\alpha$ value of the
series randomized over the rows is positive (with a standard deviation of 0.01), indicating that the
degree of freedom in selecting the order of nodes still preserves
autocorrelation in the overall signal (Figure \ref{H_random}).
Nevertheless, it is still possible that preserving also the spatial
order of the rows, that is, arraying the starting points of
each signal series according to the distance from a first node (pivot), will
affect the autocorrelation of the global signal. This may intuitively
depend on the fact that neighbors are more likely to share neighbors
than distant nodes. Neighboring nodes will thus have similar series' order, whereas for a random order of nodes, the series'order will be generated by a larger number of transpositions. For a more formal description see \cite{Lee, Aurenhammer} . We ultimately
tested this hypothesis by arraying the nodes according to the distance
from a pivot vertex and obtained a higher autocorrelation (Figure \ref{H_ordered}).
We applied DFA to spatially encoded series generated by the Gleditsch
database with the aim of assessing homogeneity of space in the WTW
and its evolution in time\cite{Gleditsch}. Figure \ref{H_non_ordered} displays the $\alpha$ value of deflated and non-deflated trade volumes, with the series composing the signal not ordered according to the
distance from one starting vertex\footnote{Missing points refer to missing values due to computational problems. Nevertheless, the trend is still
easily inferred and the time sample significant.}.  As expected, we
observe an autocorrelation in space indicating that space matters in
shaping trading relationships. The non-deflated signal seems to
exhibit a stationary trend, hinting that the role of distance has not
changed throughout time. However, if we look at the deflated signal we
can clearly spot three main periods: a growing trend that goes from
the post-war to the 1960s, a descending trend from the early 1980s on
ward and a stationary period in between. We recently investigated the
development of the spatial filling - the degree of stretchiness of the
WTW - and obtained similar results: the WTW undergone to a shrinking
phase up to the mid 1960s and an expanding phase from the mid 1970s,
though followed by a new stationary phase in the late 1990s
\cite{RuzzenentiPiccioloBasosiGarlaschelli}
.
Finally, we ordered the rows according to the spatial distance from a
starting node set as pivot. As expected, the $\alpha$ exponent of the
spatial series that preserved both, the order of neighbors for every
node and the global order of nodes, is greater than the $\alpha$
exponent that preserves just the former one, indicating a higher
autocorrelation in the signal (Figure \ref{H_ordered}). Furthemore, even in
this case, the $\alpha$ exponent seems not to exhibit any
stationarity. The same three phases, though smoothed, are still
present. Our results are consistent with results from recent analysis
based on improved methodologies of gravity models that show that,
after the late 1970s, distances became less constraining in the WTW,
hinting declining costs of transports world wide
\cite{CoeSubramanianTamirisa, Yotov, LinSim, Lin}.
We performed the DFA on a different set of data with the aim of
extending the analysis beyond the year 2000. We used the BACI data set
developed by CEPII\footnote{$http://www.cepii.fr$} and we obtained a declining trend that further
confirms previous results, suggesting that transport costs kept
decreasing after the year 2000. However, albeit we were able to
resolve the distance puzzle, we attained a new, intriguing puzzle that
we could call, mimicking the former one, \emph{mass puzzle}. BACI data
set provides trade flows in both mass unit and monetary unit
(volumes). Surprisingly, the $\alpha$ exponent of the signal
concerning trades in mass unit is lower than the signal expressing
trades in monetary unit. This result overturns our expectations that
trades in mass should be more bond to distances than trades in money.
We conjectured that bulk and heavy commodities might be hauled with
more efficient transport modes, like sea shipping, and thereby could
be less affected by the distance travelled per unit of mass.
Nevertheless, this is just a conjecture and more investigation is
needed.
\begin{figure}
\includegraphics[scale=0.3]{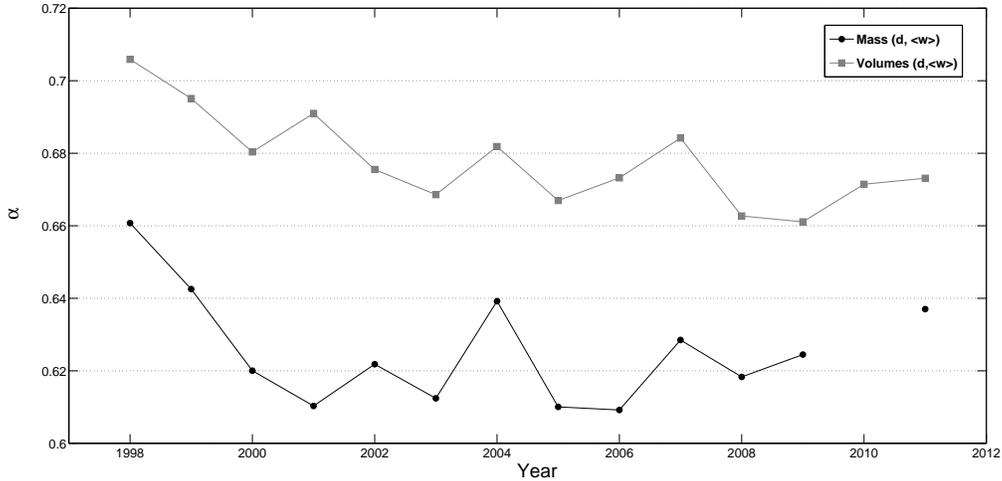}
\caption{Evolution of the $\alpha$ exponent for the WTW, deflated,
with ordered nodes, in mass and trade volumes}
\label{mass_volumes}
\end{figure}
\section{Conclusions}
\label{conc}
We proposed a method based on DFA to assess the homogeneity of space
for spatially embedded networks and tested this method on the World 
Trade Eeb. We showed that encoding the signal by preserving both the
distance from every node and the distance of every node from a pivotal
node, increases the autocorrelation in the series. Furthemore, we
showed that washing out the toplogical effects from the signal, by
means of a deflator generated with an ERG (Exponential Random Graphs)
null model, the autocorrelation further, is increasing or decreasing in
agreement with results obtained by other measures of the spatial
embeddedness of the network.  Finally, we showed that the degree of
autocorrelation decreased after the mid 1970s, indicating decreasing
\emph{cost} of distances, as expected by the \emph{distance puzzle}.
We obtained this result without adding any additional variable to the
model, merely relying on data of trade flows and their topology.
\section{Acknowledgments}
FR  acknowledges support from the FESSUD project on
``Financialisation, economy, society and sustainable development'',
Seventh Framework Programme, EU.
\appendix
\section{Null Models}
\label{appendix}
The method used to carry on the analysis of the World Trade Web
implements a recently proposed procedure \cite{SquartiniGarlaschelli}
, developed inside the exponential random graph theoretical framework
\cite{ParkNewman} . The method is composed by two main steps: the
first one is the maximization of the Shannon entropy over a previously
chosen set of graphs, $\mathcal{G}$
\begin{equation}
S=-\sum_{G\in\mathcal{G}}P(G)\ln P(G)
\end{equation}
\noindent under a number of imposed constraints
\cite{SquartiniGarlaschelli,Jaynes}, generically indicated as
\begin{equation}
\sum_{G\in\mathcal{G}}P(G)=1 \, ,  \qquad  \qquad
\sum_{G\in\mathcal{G}}P(G)\pi_{a}(G)=\langle\pi_a\rangle,\:\forall\:a
\end{equation}
\noindent (note the generality of the formalism, above: $G$ can be a
directed, undirected, binary or weighted network). We can immediately
choose the set $\mathcal{G}$ as the grandcanonical ensemble of BDNs,
i.e. the collection of networks with the same number of nodes of the
observed one (say $N$) and a number of links, $L$, varying from zero
to the maximum (i.e. $N(N-1)$). This prescription leads to the
exponential distribution over the previously chosen ensemble
\begin{equation}
P(G|\vec{\theta})=\frac{e^{-H(G,\:\vec{\theta})}}{Z(\vec{\theta})},
\end{equation}
\noindent whose coefficients are functions of the hamiltonian,
$H(G,\:\vec{\theta})=\sum_{a}\theta_{a}\pi_{a}(G)$, which is the
linear combination of the chosen constraints. The normalization
constant, $Z(\vec{\theta})\equiv\sum_{G\in\mathcal{G}}e^{-H(G,\:\vec{\theta})}$,
is the partition function \cite{SquartiniGarlaschelli, ParkNewman,
Jaynes}.
The second step prescribes how to numerically evaluate the unknown
Lagrange multipliers $\theta_{a}$. Let us consider the log-likelihood
function $\ln\mathcal{L}(\vec{\theta})=\ln P(G|\vec{\theta})$ and
maximize it with respect to the unknown parameters. In other words, we
have to find the value $\vec{\theta}^*$ of the multipliers satisfying
the system
\begin{equation}
\frac{\partial\ln\mathcal{L}(\vec{\theta})}{\partial\theta_a}\bigg|_{\vec{\theta}^*}\equiv0,\:\forall\:a,
\end{equation}
\noindent or, that is the same,
\begin{equation}
\pi_{a}(G)=\langle\pi_a\rangle(\vec{\theta}^*)\equiv\langle\pi_a\rangle^*,\:\forall\:a
\label{exp}
\end{equation}
\noindent i.e. a list of equations imposing the value of the expected
parameters to be equal to the observed one \cite{SquartiniGarlaschelli}. Note
that the term ``expected'', here, refers to the weighted average taken
on the grandcanonical ensemble, the weights being the probability
coefficients defined above.
So, once the unknown parameters have been found, it is possibile to
evaluate the expected value of any other topological quantity of
interest, $X$:
\begin{equation}
\langle X\rangle^*=\sum_{G\in\mathcal{G}}X(G)P(G|\vec{\theta}^*).
\end{equation}
Because of the difficulty to analytically calculate the expected value
of the quantities commonly used in complex networks theory, it is
often necessary to rest upon the linear approximation method: $\langle
X\rangle^*\simeq X(\langle G\rangle^*)$, where $\langle G\rangle^*$
indicates the expected adjacency matrix, whose elements are $\langle
a_{ij}\rangle^*\equiv p_{ij}^*$.
This is a very general prescription, valid for binary, weighted,
undirected or directed networks: since the WTW has been considered in
its binary, directed representation, the generic adjacency matrix $G$
will be indicated, from now on, with the usual letter $A$.
For the weighted directed version of the WTW, a very usefull null
model is the \emph{weighted directed configuration model} (WDCM), that
imposes the $in$ and $out$ strenght sequence for every node. The
Hamiltonian will thus be:
\begin{equation}
H_{WDCM}=\sum_{i}(\alpha_{i}\:s_{i}^{in}+\beta_{i}\:s_{i}^{out})=\sum_{i\neq
j}(\alpha_{j}+\beta_{i})\:w_{ij}
\end{equation}


\begin{thebibliography}{99}
\bibitem{Barthelemy} M. Barth\'elemy, {\it Spatial networks}, Physics
Reports {\bf 499} (2011), 1-101
\bibitem{Tinbergen} J. Tinbergen, {\it Shaping the World Economy:
Suggestions for an International Economic Policy}.,  Twentieth Century
Fund, New York (1962)
\bibitem{DuenasFagiolo}
M. Due\~nas and G. Fagiolo, {\it Modeling the International-Trade
Network: a gravity approach},  J. Econ. Interact. Coord., {\bf 8}
(2013) 155-178
\bibitem{GarlaschelliLoffredo2004a}
D. Garlaschelli and M. I. Loffredo, {\it Structure and evolution of
the world trade network}, Physica A, {\bf 355} (2005) 138-144
\bibitem{GarlaschelliLoffredo2004b}
D. Garlaschelli and M. I. Loffredo, {\it Fitness-dependent topological
properties of the World Trade Web}, Phys. Rev. Lett. {\bf 93} (2004)
188701
\bibitem{SquartiniFagioloGarlaschelli} T. Squartini, G. Fagiolo and D.
Garlaschelli, {\it Randomizing world trade I. A binary network
analysis}, Phys. Rev. E {\bf 84} (2011) 046117
\bibitem{AndersonYotov} J. Anderson and Y. V. Yotov, {\it Gold
Standard Gravity}, NBER Working Paper No. 17835, National Bureau of
Economic Research (February 2012)
\bibitem{BuchKleinertToubal} C. M. Buch, J. Kleinert and F. Toubal
{\it The distance puzzle: on the interpretation of the distance
coefficient in gravity equations}, Economics Letters, {\bf 83} (2004)
293-298
\bibitem{RuzzenentiBasosi} F. Ruzzenenti and R. Basosi,  {\it
Evaluation of the energy efficiency evolution in the European road
freight transport sector}, Energy Policy {\bf 37} (2009) 4079-4085
\bibitem{Stanley1} H.E. Stanley, S.V. Buldyrev, A.L. Goldberger, Z.D.
Goldberger, S. Havlin, R.N. Mantegna, S.M. Ossadnik, C.-K. Peng and M.
Simons, {\it Statistical mechanics in biology: how ubiquitous are
long-range correlations?}, Physica A {\bf 205} (1994) 214-253
\bibitem{Stanley2} K. Hu, P. Ch. Ivanov, Z. Chen, P. Carpena, H. E.
Stanley, {\it Effect of trends on detrended fluctuation analysis},
Phys. Rev. E {\bf 64} (2001) 011114 [19 pages]
\bibitem{Havlin} J. W. Kantelhardt, E. Koscielny-Bunde, H. H. Rego, S.
Havlin, A. Bunde, {\it Detecting long-range correlations with detrended
fluctuation analysis}, Physica A {\bf 295} (2001) 441-454
\bibitem{Hurst} H. E. Hurst, {\it Long-term storage capacity of reservoirs},
Trans. Amer. Soc. Civil Eng., {\bf 116} (1951), 770-808
\bibitem{Mandelbrot} B. B. Mandelbrot, {\it The fractal geometry of
nature}, W. H. Freeman and Co., New York (1983)
\bibitem{Stanley3} C.-K. Peng, S. V. Buldyrev, S. Havlin, M. Simons,
H. E. Stanley and A. L. Goldberger, {\it Mosaic organization of DNA
nucleotides}, Phys. Rev. E {\bf 49} (1994) 1685-1689
\bibitem{Stanley4} D. Horvatic, H. E. Stanley, and B. Podobnik, {\it Detrended Cross-Correlation Analysis for Non-Stationary Time Series with Periodic Trends}, Europhys. Lett. (EPL) {\bf 94} (2011) 18007 
\bibitem{PiccioloSquartiniRuzzenentiBasosiGarlaschelli} F. Picciolo,
T. Squartini, F. Ruzzenenti, R. Basosi and D. Garlaschelli, {\it The
Role of Distances in the World Trade Web}, in Conference Proceedings
``The 8th International Conference on Signal Image Technology and
Internet Based Systems" (SITIS 2012, Sorrento, Italy), Conference
Publishing Services, (2012) 784-792
\bibitem{SquartinietalBookChapter} T. Squartini, F. Picciolo, F. Ruzzenenti, R. Basosi and D. Garlaschelli, {\it Disentangling spatial and non-spatial effects in real complex networks}. In: Complex Networks and their applications. Editors: Cherifi, Hocine. (2013)Cambridge Scholars Publishing, Cambridge (UK).
\bibitem{SquartinietalWrecip} T. Squartini, F. Picciolo, F. Ruzzenenti and D. Garlaschelli, {\it Reciprocity of weighted networks}   Sci. Rep. {\bf 3}, 2729 (2013).
\bibitem{Lee} Der-Tsai Lee, {\it On k-Nearest Neighbor Voronoi
Diagrams in the Plane}, Computers, IEEE Transactions on  Computers,
{\bf C-31} (1982) 478-487
\bibitem{Aurenhammer} F. Aurenhammer, {\it Voronoi diagrams - a survey
of a fundamental geometric data structure}, ACM Computing Surveys
(CSUR) {\bf 23} (1991) 345-405
\bibitem{Gleditsch} K. S. Gleditsch, {\it Expanded Trade and GDP
Data}, Journal of Conflict Resolution {\bf 46} (2002) 712-724
\bibitem{RuzzenentiPiccioloBasosiGarlaschelli} F. Ruzzenenti, F.
Picciolo, R. Basosi and D. Garlaschelli, {\it Spatial effects in real
networks: Measures, null models, and applications}, Phys. Rev. E {\bf
86} (2012) 066110  [13 pages]
\bibitem{CoeSubramanianTamirisa} D. T. Coe, A. Subramanian and N. T.
Tamirisa, {\it The Missing Globalization Puzzle: Evidence of the
Declining Importance of Distance}, in ``International Monetary Fund
Technical reports",  IMF Staff Papers {\bf 54} (2007, International
Monetary Fund) 34-58
\bibitem{Yotov} Y.V. Yotov, {\it A simple solution to the distance
puzzle in international trade}, Economics Letters {\bf 117} (2012)
794-798
\bibitem{LinSim} F. Lin and N. C. S. Sim, {\it Death of distance and
the distance puzzle}, Economics Letters {\bf 116} (2012) 225-228
\bibitem{Lin} F. Lin, {\it Are distance effects really a puzzle?},
Economic Modelling {\bf 31} (2013) 684-689
\bibitem{SquartiniGarlaschelli} T. Squartini and D. Garlaschelli, {\it
Analytical maximum-likelihood method to detect patterns in real
networks}, New Journal of Physics {\bf 13} (2011) 083001
\bibitem{ParkNewman} J. Park and M. E. J. Newman, {\it Statistical
mechanics of networks}, Phys. Rev. E {\bf 70} (2004) 066117 [13 pages]
\bibitem{Jaynes} E. T. Jaynes, {\it Information theory and
statistical mechanics}, Phys. Rev. {\bf 106} (1957) 620-630
\end{thebibliography}
\end{document}